\documentclass[sigconf]{acmart}
\def\BibTeX{{\rm B\kern-.05em{\sc i\kern-.025em b}\kern-.08emT\kern-.1667em\lower.7ex\hbox{E}\kern-.125emX}}

\usepackage{caption}
\usepackage{array}
\usepackage{setspace} 
\usepackage{lipsum}
\usepackage{colortbl}
\usepackage{xcolor}

\usepackage{nicefrac}
\usepackage{siunitx}
\usepackage{array,framed}
\usepackage{booktabs}
\usepackage{
  color,
  float,
  epsfig,
  wrapfig,
  graphics,
  graphicx,
  subcaption
}

\usepackage{textcomp,amssymb}
\usepackage{setspace}
\usepackage{latexsym,fancyhdr,url}
\usepackage{enumerate}
\usepackage[ruled,lined]{algorithm2e}
\usepackage{algpseudocode}
\usepackage{graphics}
\usepackage{xparse} 
\usepackage{xspace}
\usepackage{multirow}
\usepackage{csvsimple}
\usepackage{balance}
\usepackage{diagbox} 

\usepackage{graphicx}
\usepackage{subcaption}

\usepackage{
  tikz,
  pgfplots,
  pgfplotstable
}

\usetikzlibrary{
  shapes.geometric,
  arrows,
  external,
  pgfplots.groupplots,
  matrix
}

\pgfplotsset{compat=1.9}

\usepackage{mathtools}

\DeclareMathAlphabet{\mathcal}{OMS}{cmsy}{m}{n}

\DeclareGraphicsExtensions{%
    .png,.PNG,%
    .pdf,.PDF,%
    .jpg,.mps,.jpeg,.jbig2,.jb2,.JPG,.JPEG,.JBIG2,.JB2}

\usepackage{xparse}
\newcommand{\bnm}{\begin{newmath}}
\newcommand{\enm}{\end{newmath}}

\newcommand{\bea}{\begin{eqnarray*}}%
\newcommand{\eea}{\end{eqnarray*}}%

\newcommand{\bne}{\begin{newequation}}
\newcommand{\ene}{\end{newequation}}

\newcommand{\bal}{\begin{newalign}}
\newcommand{\eal}{\end{newalign}}

\newenvironment{newalign}{\begin{align}%
\setlength{\abovedisplayskip}{4pt}%
\setlength{\belowdisplayskip}{4pt}%
\setlength{\abovedisplayshortskip}{6pt}%
\setlength{\belowdisplayshortskip}{6pt} }{\end{align}}

\newenvironment{newmath}{\begin{displaymath}%
\setlength{\abovedisplayskip}{4pt}%
\setlength{\belowdisplayskip}{4pt}%
\setlength{\abovedisplayshortskip}{6pt}%
\setlength{\belowdisplayshortskip}{6pt} }{\end{displaymath}}

\newenvironment{newequation}{\begin{equation}%
\setlength{\abovedisplayskip}{4pt}%
\setlength{\belowdisplayskip}{4pt}%
\setlength{\abovedisplayshortskip}{6pt}%
\setlength{\belowdisplayshortskip}{6pt} }{\end{equation}}

\newcounter{ctr}

\newcounter{mytable}
\def\mytable{\begin{centering}\refstepcounter{mytable}}
\def\endmytable{\end{centering}}

\newcounter{myfig}
\def\myfig{\begin{centering}\refstepcounter{myfig}}
\def\endmyfig{\end{centering}}

\newlength{\saveparindent}
\newlength{\saveparskip}
\newcommand{\E}{{\rm I\kern-.3em E}}

\renewcommand{\eqref}[1]{\mbox{Equation~(\ref{#1})}}

\def \part {part}

\renewcommand{\paragraph}[1]{\vspace*{6pt}\noindent\textbf{#1}\;}

\def \blackslug{\hbox{\hskip 1pt \vrule width 4pt height 8pt
    depth 1.5pt \hskip 1pt}}
\def \qed{\quad\blackslug\lower 8.5pt\null\par}

\newcounter{mynote}[section]

\newcommand\ignore[1]{}

\newcounter{rcnote}[section]

\newcounter{mrnote}[section]

\newcounter{fknote}[section]

\newcounter{anote}[section]

\DeclareMathSymbol{\mlq}{\mathord}{operators}{``}
\DeclareMathSymbol{\mrq}{\mathord}{operators}{`'}

\newcommand{\rhf}[2]{R_{f, \gamma}}

\DeclareDocumentCommand{\edist}{o o}{
  \ensuremath{
    \IfNoValueTF{#1}{{d}}{{\sf d}(#1,#2)}
  }
}

\newcommand{\olrk}[1]{\ifx\nursymbol#1\else\!\!\mskip4.5mu plus 0.5mu\left(\mskip0.5mu plus0.5mu #1\mskip1.5mu plus0.5mu \right)\fi}

\NewDocumentCommand{\indseq}{ O{1} O{r} }{{#1}\ldots {#2}}

\usepackage{hyperref}

\definecolor{myblue}{RGB}{0, 122, 204}
\definecolor{myred}{RGB}{255, 0, 0}
\definecolor{mygreen}{RGB}{0, 153, 51}
\definecolor{mygray}{RGB}{240, 240, 240}

\hypersetup{
    colorlinks=true,
    linkcolor=myblue,     
    citecolor=mygreen,    
    urlcolor=myred,       
    linkbordercolor=myblue, 
}

\begin{document}
\fancyhead{}
\def\thetitle{Hide in Plain Sight: Clean-Label Backdoor for Auditing Membership Inference}
\title{\thetitle}

\author{Depeng Chen, Hao Chen, Hulin Jin, Jie Cui, Hong Zhong }
\affiliation{\small{School of Computer Science, Anhui University}
}

\date{}

\begin{abstract}
Membership inference attacks (MIAs) are critical tools for assessing privacy risks and ensuring compliance with regulations like the General Data Protection Regulation (GDPR). However, their potential for auditing unauthorized use of data remains underexplored. To bridge this gap, we propose a novel clean-label backdoor-based approach for MIAs, designed specifically for robust and stealthy data auditing. Unlike conventional methods that rely on detectable poisoned samples with altered labels, our approach retains natural labels, enhancing stealthiness even at low poisoning rates.

Our approach employs an optimal trigger generated by a shadow model that mimics the target model’s behavior. This design minimizes the feature-space distance between triggered samples and the source class while preserving the original data labels. The result is a powerful and undetectable auditing mechanism that overcomes limitations of existing approaches, such as label inconsistencies and visual artifacts in poisoned samples.

The proposed method enables robust data auditing through black-box access, achieving high attack success rates across diverse datasets and model architectures. Additionally, it addresses challenges related to trigger stealthiness and poisoning durability, establishing itself as a practical and effective solution for data auditing. Comprehensive experiments validate the efficacy and generalizability of our approach, outperforming several baseline methods in both stealth and attack success metrics.
\end{abstract}

\maketitle


\section{Introduction}
\label{sec:intro}
With the advancement of contemporary technology, deep learning has achieved remarkable success across various domains, including image recognition \cite{he2023toward}, face recognition \cite{face}, medical image analysis \cite{cancer}, and natural language processing \cite{speech}. High-quality, reliable real-world data is the cornerstone of these deep learning advancements. However, this data often contains significant amounts of user privacy information, raising concerns about unauthorized use. For example, a British hospital shared the data of 1.6 million patients with the artificial intelligence company DeepMind without obtaining patient consent \cite{deep21}. Such actions represent severe breaches of privacy and violate laws like the GDPR. 

Despite these growing concerns, users often struggle to determine whether their data has been illicitly collected and used for model training, as data can be easily harvested from the web and social media without explicit authorization. This critical issue remains insufficiently addressed in current machine learning research, highlighting the need for effective mechanisms to audit data usage. Our work aims to fill this gap by proposing novel methods that empower users to detect unauthorized data use in model training.

\textbf{Membership inference attacks} (MIAs) are a relatively recent tactic in which attackers determine whether a specific data sample was used to train a model \cite{shokri2017membership} \cite{salem2019ml} \cite{liu2022membership}. These attacks serve two primary purposes: assessing the privacy risks of a target model \cite{zhang2021membership} \cite{yeom2018privacy} and auditing data usage to identify potential breaches of regulations like GDPR \cite{hu2022membership}. In this study, we utilize MIAs to help users detect unauthorized use of their data. However, existing MIA methods often fall short in addressing this issue.

\begin{figure*}[t]  
  \centering
  \includegraphics[width=\textwidth]{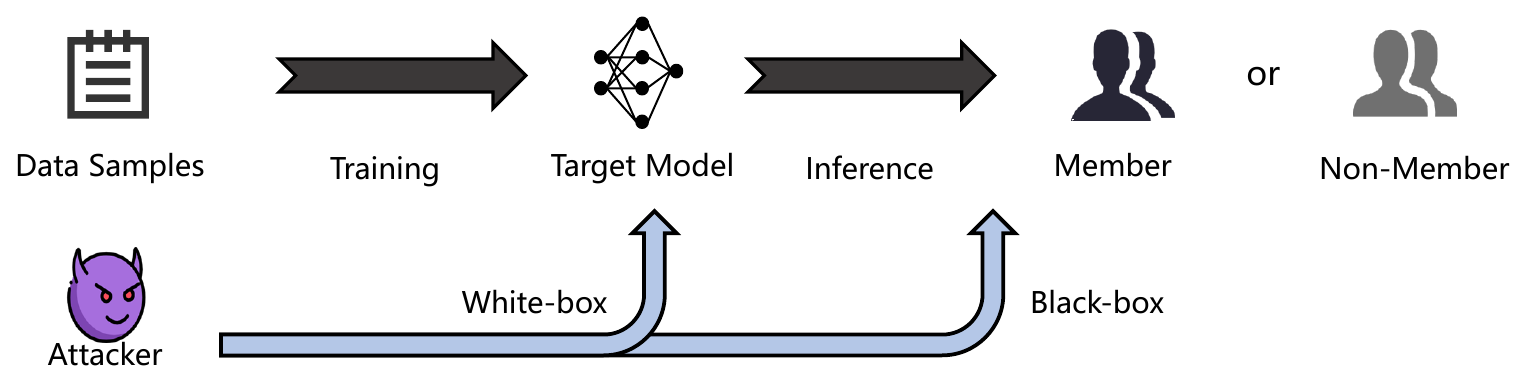}
    \caption{The process of membership inference attack under black-box and white-box attacks. This illustration highlights the differences in attack methods based on the level of knowledge available to the attacker: black-box attacks rely on the model’s outputs, while white-box attacks utilize full access to the model’s internals.}
  \label{fig:MIA}
\end{figure*}

The effectiveness of MIAs hinges on exploiting the tendency of machine learning models to overfit, as models typically exhibit higher confidence in the samples they were trained on, which is reflected in their output scores \cite{chen2022amplifying} \cite{yeom2018privacy}. For instance, an attacker can leverage output scores from a shadow model to train a binary attack model, identifying whether clinical records were used in training a model related to a specific ailment—thereby violating privacy. Figure \ref{fig:MIA} illustrates the typical MIA process used to assess privacy leakage. Despite their focus on privacy risks, current MIA methods rarely explore their potential for data auditing.

Current membership inference methods often face limitations when the target model demonstrates strong generalization and only provides predicted labels without output probabilities \cite{jia2019memguard}. In such scenarios, these methods struggle to achieve the necessary accuracy. To overcome this challenge, Hu et al. \cite{hu2022membership} introduced the use of backdoor attacks in MIAs as a novel approach to data auditing. By embedding triggers in tainted samples, users can determine whether their data was used in the training process: a correct prediction suggests the sample was not part of the training set, while a matching target label indicates it was \cite{gu2019badnets}\cite{li2024backdoor}.   

However, in backdoor-based membership inference attack (MIA) techniques, achieving a high success rate in identifying members often comes at the expense of the model's utility \cite{li2024backdoor}. A significant drawback is the inconsistency between the labels and the poisoned samples containing the backdoor \cite{li2024backdoor}, which makes them easily detectable by inspectors and difficult to deploy in real-world scenarios.

To address these limitations, we propose a novel MIA method that utilizes clean-label backdoor strategies. This approach mitigates the challenges of existing backdoor methods by generating poisoned samples that closely resemble the original data, thereby enhancing concealment. Through extensive experimentation across diverse datasets and deep neural network architectures, we demonstrate the effectiveness of our proposed method, particularly at low poisoning rates. Additionally, we investigate various factors that influence the outcomes of our experiments.
\textbf{Our contributions can be summarized as follows:}
\begin{itemize}
    \item We propose a novel method for performing membership inference attacks, called "clean label backdoor membership inference". This method ensures that the labels of data samples remain consistent without altering them. By adopting this technique, we overcome the limitation of existing membership inference methods that rely on backdoor attacks, where poisoned samples are easily detectable. We enhance the concealment of poisoned samples by carefully designing subtle triggers that closely resemble the original data.
    
    \item Our attack method surpass the limitations of current backdoor attack techniques that require knowledge of real target model training samples. Instead, we use shadow models and shadow datasets to design the triggers. In addition, even in the case of extremely low poisoning rates, our method can still achieve high attack performance.
    
    \item Additionally, our approach relaxes the requirement that the shadow model must have the same architecture as the target model. We conduct extensive experimental analysis to investigate various factors influencing the attack performance and analyze the underlying causes behind the attack's success.
    
    \item Finally, we provide a comparative analysis of our attack method against existing membership inference attack techniques that use backdoors. Experimental results demonstrate that, despite weaker assumptions about the adversary's knowledge, our approach consistently outperforms existing attacks while minimizing the impact on model utility. We validate the effectiveness of our proposed method across various public datasets and models.
\end{itemize}



\section{PRELIMINARIES }
\label{sec:relwork}
This work focuses on the application of supervised machine learning, specifically using membership inference attacks (MIAs) to audit user privacy. In this section, we outline the key concepts of MIAs and define the threat model central to our approach.

\begin{figure*}[t]  
  \centering
  \includegraphics[width=\textwidth]{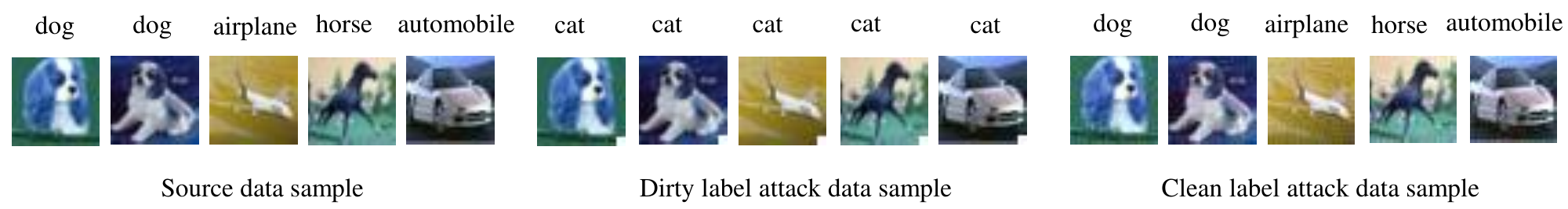}
  \caption{\textbf{ Examples of poisoned samples for the target class "cat". Left: Original sample.  Middle: Dirty-label attack with a visible trigger (white square in the bottom right corner). Right: clean-label attack data sample Clean-label attack sample with perturbation limit $\epsilon$ = 16/255. We provide the label of each sample on the top. For the clean-label poisoning samples, they look like natural ones. Our approach overcomes the drawbacks of easy discovery of triggers and inconsistent labels.}}
  \label{fig:compare}
\end{figure*}

\subsection{Membership Inference Attacks in Machine Learning }
\label{sec:overview1}
In this work, our focus is on supervised classification tasks within the realm of computer vision. Specifically, we consider a training dataset composed of \( N \) samples, which can be 
denoted as \( D_{train} = \{ (x_{1}, y_{1}), (x_{2}, y_{2}), \ldots, (x_{N}, y_{N}) \} \). This dataset consists of both clean data \( D_{clean} \) and poisoned data \( D_{poisoned} \), such that \( D_{train} = D_{clean} \cup D_{poisoned} \). Each sample includes feature vectors \( x_i \in X \) and corresponding labels \( y_i \in Y \). The learning process involves training a deep classifier \( f(x; \theta) \) with model parameters \( \theta \) to minimize a predefined loss function \( \mathcal{L}(f(x; \theta), y) \) on the training dataset \( D_{train} \). The objective is to find the optimal parameters \( \theta^{\ast} \) by solving the following optimization problem:

\begin{equation}
    \theta^{\ast} = \underset{\theta}{\arg\min} \sum_{i=1}^{N} \mathcal{L}(f(x_i; \theta), y_i)
    \label{base}
\end{equation}

Here, \( f(x; \theta) \) refers to a deep neural network, and \( N \) represents the total number of samples in the training dataset. The loss function \( \mathcal{L} \) guides the training process. After training, the model \( f(x; \theta^{\ast}) \) is utilized to predict the labels of test samples.

Membership inference attacks (MIAs) aim to determine whether a specific data sample is part of the training set used by a target model. Formally, given a data sample \( x \), a trained machine learning model \( M \), and the external knowledge possessed by the adversary (denoted as \( K \)), a membership inference attack \( \mathcal{A} \) can be defined as a function:

\begin{equation}
\mathcal{A} : (x, M, K) \longrightarrow \{ 0, 1 \}
\end{equation}

In this context, \( x \) represents the data sample, \( M \) signifies the machine learning model, and \( K \) denotes the adversary's external knowledge. The output of the function indicates membership status, where 1 signifies that the sample is part of the training set (member), and 0 indicates that it is not part of the training set (non-member).

\subsection{Threat Model }
\label{overview1}
\textbf{Attack Goals.} In this article, we consider the user as the attacker. Our primary objective is to enable users to review their data and ascertain whether it has been illicitly utilized by unauthorized organizations. Concurrently, the attackers strive to ensure that the poisoned samples remain concealed and difficult to detect. Our secondary objective is to minimize the impact of the poisoned samples on model performance. Moreover, we assert that the poisoned samples should be visually indistinguishable from clean samples, rendering them imperceptible to human reviewers.

\textbf{Attacker Capabilities.} We aim to develop a membership inference approach that empowers data owners to discern whether their data was used in training a target model. We operate under the following assumptions:

\textbf{(1) Limited Access:} Attackers can only access the target model through a black-box method \cite{chen2022amplifying}, which excludes access to internal model information such as gradients and losses. This setup makes it more challenging for attackers.

\textbf{(2) Shadow Dataset:} The attacker possesses a shadow dataset $D_{shadow}$, which is a common setting in membership inference attack scenarios \cite{liu2022membership,salem2019ml,hu2022membership}.The attacker possesses a shadow dataset $D_{shadow}$ for crafting triggers, comprising clean samples drawn from the same distribution as $D_{clean}$. The attacker can train a shadow model \( M_{s} \) to replicate the behavior of the target model and strategically inject poisoned samples into the training set of the target model. This assumption aligns with common practices in membership inference attacks.

In subsequent ablation experiments, we will relax the assumption regarding the shadow model. Specifically, the shadow model will not necessarily need to share the same architecture as the target model, yet our method will still achieve significant efficacy in conducting attacks.

\textbf{Attack Setup and Assumptions}. In this work, we focus on leveraging clean-label backdoor attacks for membership inference, offering data owners a practical approach to auditing their data. To enhance the attack's effectiveness, we propose optimizing the trigger through the use of a shadow model. A common assumption is that the user can collect a small shadow dataset sampled from the same distribution as the target dataset \cite{wen2024membership}. Typically, this shadow dataset is built from publicly available data that shares similar distributional characteristics with the target dataset, allowing the auditing user to construct a model that closely mimics the behavior of the target model. Furthermore, our method ensures that the backdoor trigger remains difficult to detect, requiring only a low poisoning rate while still achieving a high success rate in membership inference attacks.

\begin{figure*}[t]  
  \centering
  \includegraphics[width=\textwidth]{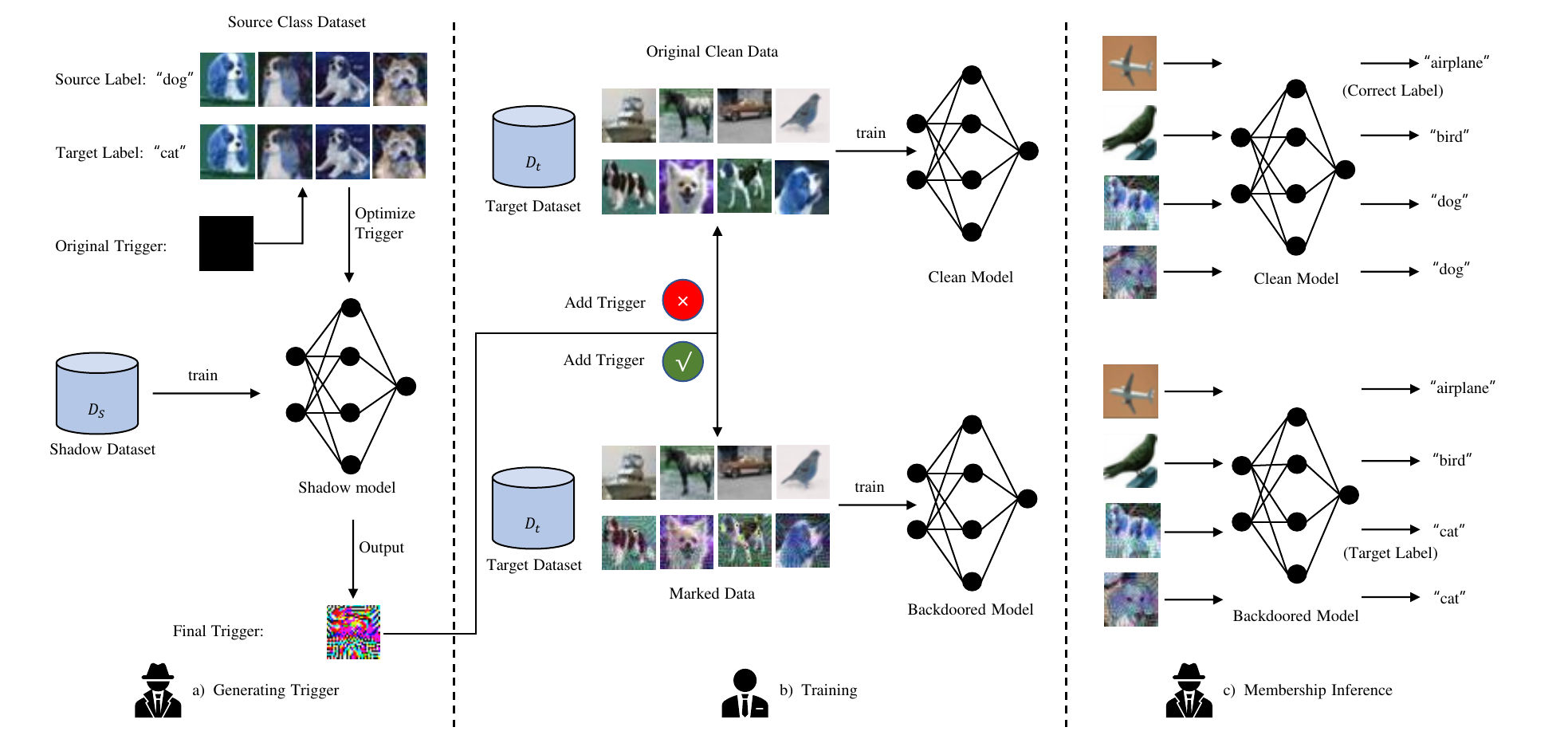}
  \caption{\textbf{An illustration of membership inference via a clean-label backdoor approach. The source label is "dog", the target label is "cat", and the trigger is trained by the shadow model. The attack method can be divided into three stages: a) The data owner trains the shadow model and uses the shadow model to train the trigger; b) Model training. Inject poisoned samples into the training set for model training; c) The data owner performs membership inference by querying the target model.}}
  \label{fig:large_image}
\end{figure*}

\section{Methodology}
\label{sec:methodology}

\subsection{Problem Statement}
\label{overview1}

Our objective is to provide users with a method to audit their data and determine whether it has been illicitly acquired and used for training machine learning models. The problem is outlined as follows:

Suppose you are a user seeking to determine if your data has been surreptitiously collected by an unauthorized organization for training machine learning models. You possess multiple data samples, denoted as $\left ( x_{1} ,y_{1} \right ), \cdots ,\left ( x_{n} ,y_{n} \right )$, some of which may contain triggers. Each data sample comprises features $x~\in ~ X$ and labels $y~\in ~ Y$, where both labels and features are consistent. Illicit organizations can obtain these data samples from websites or social media platforms, integrate them into the training dataset $D_{train}$, and proceed to train a classification model $f\left(\theta\right)$. Once the model is trained, these organizations may exploit the model for commercial purposes without user consent.

The goal of this work is to develop a membership inference method using clean-label backdoor attacks, allowing users to detect if their private data has been used by unauthorized parties for training target models. Importantly, our method will ensure that the model's utility remains intact, while the poisoned samples used in the attack will be indistinguishable from clean samples, making them difficult to detect. This provides a robust approach for users to protect their data privacy without compromising model performance.


\subsection{Design Intuition}
The main challenge in performing auditing membership inference using backdoor triggers lies in concealing the poisoned samples while maintaining their effectiveness. Traditional backdoor attacks, which modify both the appearance and label of samples, face detection risks due to label inconsistency and conspicuous triggers. Our intuition stems from addressing these limitations by maintaining label consistency and embedding imperceptible triggers within the data.

The core idea is to exploit the target model's sensitivity to subtle, carefully crafted perturbations that do not visually alter the data but can still influence the model's decision-making. By employing clean-label attacks, we ensure that the poisoned samples closely resemble their clean counterparts, making them difficult to detect through visual inspection or automated filters. This approach leverages the fact that even minor perturbations in the feature space can significantly influence a model's output.

Furthermore, by optimizing triggers through shadow models, which mimic the target model’s behavior, we can reduce the distance between the poisoned samples and their target class in the feature space. This enhances the effectiveness of our attacks without compromising the natural appearance of the samples. Our method is designed to function effectively even under low poisoning rates, ensuring minimal impact on model utility while maintaining high attack success rates.

\textbf{Clean samples.} As shown in the sample on the far right in Figure 2, in this paper, the clean data sample with a trigger proposed by us refers to a data sample that shows no abnormality in appearance and has the correct label category. This is different from the clean samples proposed in \cite{gu2019badnets}\cite{hu2022membership}. Their clean samples either have incorrect label categories or the trigger features are easily noticeable in appearance. In addition, the perturbation size of our trigger is set the same as that in \cite{zeng2023narcissus} \cite{souri2022sleeper}etc.

In essence, our clean-label strategy balances the need for attack performance with the requirement for undetectability, making it suitable for practical deployment in membership inference scenarios.

\subsection{Attack Methodology }
\label{overview1}
The current approach to auditing membership inference through backdoor methods faces challenges in real-world scenarios. Traditional techniques, such as BadNets \cite{gu2019badnets}, involve adding patches to fixed positions in data samples and altering their labels to match the target model. However, a critical limitation of this method is that the labels of poisoned samples often become inconsistent with their characteristics, making them easily detectable. 

To address this issue, we propose the use of clean labels. Figure \ref{fig:compare} illustrates the difference between clean-label and dirty label attacks relative to the source data samples. The poisoned samples produced by our clean-label poisoning method appear more natural to the human eye, making them more difficult to detect through manual review. In contrast, the method introduced by Hu et al. \cite{hu2022membership} involves adding visible white squares as triggers and forcibly setting the labels of the source data samples to the target labels. This alteration makes the poisoned samples easy to detect and filter out. Our approach does not modify the sample labels and instead sets the perturbation range to ensure that the trigger remains undetectable. Importantly, our attack method maintains high performance even with an extremely low poisoning rate, highlighting its effectiveness.

In clean-label attacks, the trigger remains imperceptible to the naked eye, while the label of the poisoned sample stays consistent with its observable characteristics. Although the straightforward approach of adding triggers without altering labels may seem effective, it often results in a notably low attack success rate. To address this, we introduce an efficient "one-to-one" clean-label attack method. \textbf{The key aspect of our method involves optimizing the trigger iteratively using a pre-trained shadow model.} This shadow model retains knowledge of the features from the same distribution dataset, allowing it to effectively train the trigger to incorporate characteristics of the target class. By adding these triggers, we reduce the distance between source and target class samples in the feature space. As a result, the poisoned sample with the added trigger maintains a natural appearance, consistent with the original sample, while becoming closer to the source class in the feature space, all without changing its original label.

Our objective is to utilize clean-label backdoor attacks for auditing membership inference, optimizing triggers using shadow models and datasets. Figure \ref{fig:large_image} illustrates the entire process of our proposed membership inference method, which consists of three key stages: a) trigger generation; b) trigger injection and deep neural network (DNN) model training; c) data owner querying the target model to detect whether its data was used for training. We will elaborate on each of these stages in further detail.

\paragraph{a) Generating Triggers}

In the first stage, the attacker, acting as the data owner, possesses the capability to generate poisoned samples. We divide trigger generation into three incremental steps to achieve this: 

\begin{itemize}
    \item  The data owner initially employs a shadow dataset \(D_{shadow}\) to train the shadow model \(M_{s}\) and initializes a trigger with the same dimensions as the data sample (initial pixel values are all set to 0). 
    \item The data owner segregates the source class dataset \(D_{source}\) from the shadow dataset, modifies its label from the source label to the target label, and incorporates the trigger into it.
    \item  With the parameters of the shadow model \(M_{s}\) fixed, the trigger is further optimized using the source class dataset, which now contains the added trigger, to produce the final optimized version.
\end{itemize}

To ensure that the poisoned samples appear more natural, \(L_{\infty}\) norm constraints (we set \(\varepsilon = 16/255\) in the experiment) are applied to the poisoned samples to add constraints. Equation \ref{control_equation} outlines the constraint conditions, while Algorithm \ref{alg:example} provides comprehensive details for trigger generation.

\begin{equation}
    \left | \left |~ x_{p} -x ~\right |  \right |_{\infty } ~\le ~ \varepsilon  \label{control_equation}
\end{equation}
where $x$ represents the original sample, $x_{p}$ denotes the poisoned sample after adding the trigger, and $\varepsilon$ signifies the constraint condition. The feature distance between the original sample and the poisoned sample is constrained by utilizing the $L_{\infty }$ norm.

\paragraph{b) Training Target Model}

Once the data owner obtains the optimized trigger from the previous stage, they select a set of data samples from the target class within the clean dataset \(D_{train}\). The trigger is then integrated into these samples, and the resulting poisoned samples are introduced into the original clean dataset \(D_{train}\). It’s important to note that when data owners publicly share their data on social media or websites, unauthorized parties can covertly collect this data without the owners' knowledge. Subsequently, these unauthorized parties use the collected data to train the target model, providing a black-box API of the trained model to end users seeking to employ it for prediction and classification tasks, often for commercial purposes.

\paragraph{c) Membership Inference}

During the membership inference attack phase, when an unauthorized party employs the collected data to train the target model, the model is backdoored. To determine whether their own data samples are included in the training set of the target model, a data owner utilizes poisoned samples to test the model and observes the divergent behaviors between the target model and a clean model. This discrepancy allows them to infer the membership of their data. To test a given input sample \(x_{test}\), the trigger is augmented by a certain ratio (e.g., three times), and empirical results demonstrate that amplifying the trigger can effectively enhance attack performance. Prior research~\cite{turner2019label} has highlighted the significance of trigger amplification during the testing phase, as unauthorized parties rarely inspect test samples compared to training data samples.

To provide statistical confidence in the membership inference results, we employ a statistical test proposed in the literature \cite{hu2022membership} that estimates the confidence level for assessing the presence of a backdoor in the target model. We define the null hypothesis \(H_0\) and the alternative hypothesis \(H_1\) as follows:

\begin{equation}
    \begin{aligned}
    \mathcal{H}_{0} & : P_{r} ~(f(x_{p}) = y_{t}~) \le \beta \\
    \mathcal{H}_{1} & : P_{r} ~(f(x_{p}) = y_{t}~) >  \beta  
    \end{aligned}  
    \label{hypothes}
\end{equation}
where \(P_{r} (f(x_{p}) = y_{t})\) represents the attack success probability of the model containing the backdoor predicting the poisoned sample as belonging to the target class, and \(\beta\) denotes the success threshold for backdoor attacks in the clean model. In this work, we set \(\beta = \frac{1}{k}\), where \(k\) represents the number of classes in the classification task, reflecting a random probability. Furthermore, we will empirically demonstrate in subsequent experiments that the backdoor attack success rate of the target model is significantly lower than the threshold \(\beta\). 

Additionally, we employ the t-test method proposed by Hu et al.~\cite{hu2022membership} to evaluate hypotheses under specified conditions. This method clarifies when the data owner can reject the null hypothesis \(H_{0}\) with a confidence level of \(1 - \tau\) by making a limited number of queries to the target model. It provides a rigorous mathematical formula and proof, expressed as follows:

\begin{equation}
    \sqrt{q-1} ~\cdot ~(\alpha -\beta )~ - \sqrt{\alpha -\alpha ^{2} }~\cdot ~ t_{\tau } > 0
    \label{compare}
\end{equation}
Here, $q$ represents the number of queries requested by the data owner, $\alpha$ denotes the attack success rate, \( \beta = \frac{1}{k} \) (where \( k \) is the total number of distinct classes in the target model), and $t_{\tau }$ stands for the $\tau$ quantile of the $t$ distribution with $q-1$ degrees of freedom. We set the number of queries to 30 and the significance level to 0.05. When the data owner queries the target model $m$ times and the attack success rate exceeds the threshold, they can assert that an unauthorized party has clandestinely utilized their own data to train the model. That is, when our attack rate exceeds the threshold, we can claim that our user data is being used privately.

\SetAlgoNlRelativeSize{0}
\LinesNumbered
\begin{algorithm}
\SetKwData{Left}{left}
\SetKwData{This}{this}
\SetKwData{Up}{up}
\SetKwFunction{Union}{Union}
\SetKwFunction{FindCompress}{FindCompress}
\SetKwInOut{Input}{Input}
\SetKwInOut{Output}{Output}
\caption{Generating Triggers}
\label{alg:example}
\Input{ Shadow model $f\left ( \theta  \right )$, Source class data samples $D_{\hat{s} }$, Shadow dataset $D_{s}$, Target label $y_{t}$, Gradient descentable trigger $\mathcal{T}_{1}$, Number of epochs $E$}
\Output{Final trigger $\mathcal{T}$}
\BlankLine
Train the shadow model $f\left ( \theta  \right )$ on the shadow dataset $D_{s}$\;
Fix the parameters of the trained shadow model $f\left ( \theta  \right )$\;
\textbf{Initialize:} $\mathcal{T}_{1}\gets 0^{d\times d} $,$~~E=  1000,~~g_{1}=  1$\;
\textbf{for each} sample $(X, Y)$ in $D_{\hat{s} }${

\hspace{0.5cm} $Y \gets y_{t}$\;
}
\For{$i\leftarrow 1$ \KwTo E}{
\BlankLine
$\mathcal{T}_{i+ 1 }\gets \mathcal{T}_{i}- \alpha \sum _{\left ( x,y_{t} \right )\in D_{\hat{s}}} 
\nabla _{\mathcal{T}_{i} } \mathcal{L}\left  ( f\left ( x +\mathcal{T}_{i};\theta \right ) ,y_{t}\right ) $\;
\BlankLine
$\mathcal{T}_{i+ 1 }\gets Proj_{\bigtriangleup } \left ( \mathcal{T}_{i+ 1 } \right ) $\hspace{0.3cm}/*Constraint Trigger */\;
\BlankLine
$g_{i}\gets \bigtriangledown   \left ( \mathcal{T}_{i+ 1 } \right ) $\hspace{0.3cm}/*Compute Gradient */\;
\BlankLine

\If{$g_{i} = 0 $ or $ i \ge   E$}{
break\;
}
}
\textbf{Return} Final trigger $\mathcal{T}$
\label{algo_disjdecomp}
\end{algorithm}

    

\section{Evaluation}
\label{sec:eval}

\subsection{Experimental Setup }
\label{over1view1}
\paragraph{Datasets.} In our experiments, we divide each dataset into three parts: the shadow dataset \(D_{shadow}\), used to train the optimization trigger; the clean training dataset \(D_{clean}\), employed for training the target model; and the model test dataset \(D_{test}\). To ensure fairness, we maintain \(\left | D_{shadow} \right | = \left | D_{clean} \right |\), with both the shadow dataset and the clean training dataset sharing the same distribution but exhibiting non-overlapping characteristics. A summary of the data descriptions and experimental parameter settings is presented in Table \ref{dataset}. We conducted experiments on three datasets to evaluate our approach, with specific details outlined as follows:

\begin{itemize}
    \item \textbf{CIFAR-10 \cite{krizhevsky2009learning}}: The CIFAR-10 dataset comprises 32×32×3 color images categorized into 10 classes, with 6,000 images per category. It contains a total of 50,000 training images and 10,000 testing images.
    \item \textbf{CIFAR-100 \cite{krizhevsky2009learning}}: The CIFAR-100 dataset is an extension of CIFAR-10, featuring 100 classes organized into 20 superclasses, with each superclass further divided into 5 classes. Each class contains 600 color images sized 32×32×3, with 500 images allocated for training and 100 for testing.
    \item \textbf{Tiny-ImageNet \cite{le2015tiny}}: The Tiny-ImageNet dataset is a subset of the larger ImageNet dataset, primarily used for image classification tasks. Each sample consists of a 64×64×3 color image, and the dataset comprises 200 categories. For each category, there are distinct training, validation, and test sets, specifically including 500 training samples, 50 validation samples, and 50 test samples.

\end{itemize}

\begin{table}
  \centering
  \caption{\textbf{Datasets description and experiment parameter settings}}
  \renewcommand{\arraystretch}{1.1}
  \captionsetup{size=large  } 
  \resizebox{\columnwidth}{!}{
  \begin{tabular}{>{\raggedright}m{2.4cm}c c c c c c }
    \toprule
    Dataset &  CIFAR-10 & CIFAR-100 & Tiny-ImageNet \\
    \midrule
    \# Classes & 10 & 100  & 200  \\
    Sample Shape & (32$\times 32\times$3) & (32$\times 32\times$3)  & (64$\times 64\times$3)  \\
    Total Samples & 25000 & 25000  & 50000  \\
    Poison Ratio & 0.1\% & 0.1\% & 0.1\%  \\
    Poison Number & 25 & 25 & 50  \\
    Source Class & 3(Cat) & 3(Bear) & 3(Tailed frog)  \\
    Target Class & 2(Bird) & 2(Baby) & 2(Bullfrog)  \\
    Threshold & 23.35\% & 10.79\% & 9.94\%  \\
    
    \bottomrule
  \end{tabular}
  }
  \label{dataset}
\end{table}

\begin{table*}[t]
  \centering
  \caption{\textbf{Experimental results and comparison of CIFAR-10, CIFAR-100 and Tiny-ImageNet. BadNets-c and Ours represent clean label poisoning, BadNets-d and MIB\cite{hu2022membership} represent dirty label poisoning. Among them, red marks the best ASR. The poisoning rate is $0.1\%$. SA is the abbreviation of Sleeper Agent method.The clean method refers to attacking a clean model without backdoors, and this method is used as our baseline.}}
  \renewcommand{\arraystretch}{1.0}
  \resizebox{0.9\textwidth}{!}{%
    \begin{tabular}{ccccccccccc}  
      \hline
      & & \multicolumn{3}{c}{CIFAR-10} & \multicolumn{3}{c}{CIFAR-100} & \multicolumn{3}{c}{Tiny-ImageNet}\\  
      \cline{3-11}
      & & Test & CAD & ASR & Test & CAD & ASR & Test & CAD & ASR\\
      \hline
      \multicolumn{2}{>{\raggedright}m{2.0cm}}{\textbf{ResNet18}} &  &  &  &  &  &  &  &  & \\
      & \multicolumn{1}{>{\raggedright}m{1.6cm}}{Clean} & 92.31 $\pm$ 0.22 & 0 & 0.41 $\pm$ 0.12 & 67.54 $\pm$ 0.09 & 0 & 0.81 $\pm$ 0.31 & 48.90 $\pm$ 0.04 & 0 & 0.66 $\pm$ 0.52 \\
      & \multicolumn{1}{>{\raggedright}m{1.6cm}}{BadNet-c\cite{gu2019badnets}} & 88.32 $\pm$ 0.03 & -2.99 & 2.18 $\pm$ 0.14 & 66.68 $\pm$ 0.19 & -0.86 & 0.65 $\pm$ 0.07 & 48.67 $\pm$ 0.08 & -0.23 & 0.27 $\pm$ 0.03 \\
      & \multicolumn{1}{>{\raggedright}m{1.6cm}}{BadNet-d\cite{gu2019badnets}} & 89.22 $\pm$ 0.06 & -3.09 & 55.54 $\pm$ 0.19 & 65.59 $\pm$ 0.14 & -1.95 & 2.11 $\pm$ 0.34 & 48.37 $\pm$ 0.11 & -0.53 & 78.47 $\pm$ 0.32 \\
      & \multicolumn{1}{>{\raggedright}m{1.6cm}}{MIB\cite{hu2022membership}} & 90.27 $\pm$ 0.21 & -2.04 & 3.45 $\pm$ 0.32 & 64.79 $\pm$ 0.11 & -2.75 & 2.17 $\pm$ 0.33 & 47.16 $\pm$ 0.09 & -1.74 & 0.74 $\pm$ 0.13 \\
      & \multicolumn{1}{>{\raggedright}m{1.6cm}}{SA\cite{souri2022sleeper}} & 91.70 $\pm$ 0.32 & -0.61 & 42.00 $\pm$ 1.41 & 66.15 $\pm$ 0.16 & -1.39 & \cellcolor{red!25}79.67 $\pm$ 0.64 & 47.38 $\pm$ 0.13 & -1.52 & 46.65 $\pm$ 0.78 \\
      & \multicolumn{1}{>{\raggedright}m{1.6cm}}{Ours} & 91.83 $\pm$ 0.32 & -0.48 & \cellcolor{red!25}60.79 $\pm$ 1.41 & 67.15 $\pm$ 0.16 & -0.39 & 69.67 $\pm$ 0.64 & 47.28 $\pm$ 0.13 & -1.62 & \cellcolor{red!25}82.65 $\pm$ 0.78 \\
      \hline

      \multicolumn{2}{>{\raggedright}m{2.0cm}}{\textbf{VGG16}} &  &  &  &  &  &  &  &  & \\
      & \multicolumn{1}{>{\raggedright}m{1.6cm}}{Clean} & 88.76 $\pm$ 0.19 & 0 & 1.34 $\pm$ 0.29 & 65.45 $\pm$ 0.07 & 0 & 1.21 $\pm$ 0.23 & 57.82 $\pm$ 0.06 & 0 & 0.35 $\pm$ 0.12 \\
      & \multicolumn{1}{>{\raggedright}m{1.6cm}}{BadNet-c\cite{gu2019badnets}} & 86.03 $\pm$ 0.03 & -2.73 & 2.98 $\pm$ 0.04 & 65.05 $\pm$ 0.06 & -0.40 & 1.09 $\pm$ 0.02 & 57.62 $\pm$ 0.06 & -0.20 & 0.42 $\pm$ 0.25 \\
      & \multicolumn{1}{>{\raggedright}m{1.6cm}}{BadNet-d\cite{gu2019badnets}} & 87.73 $\pm$ 0.06 & -1.03 & 7.28 $\pm$ 0.09 & 64.81 $\pm$ 0.14 & -0.64 & 3.14 $\pm$ 0.08 & 57.52 $\pm$ 0.08 & -0.30 & 48.34 $\pm$ 0.29 \\
      & \multicolumn{1}{>{\raggedright}m{1.6cm}}{MIB\cite{hu2022membership}} & 87.88 $\pm$ 0.25 & -0.88 & 5.27 $\pm$ 0.18 & 65.34 $\pm$ 0.14 & -0.11 & 1.89 $\pm$ 0.11 & 57.31 $\pm$ 0.13 & -0.51 & 6.42 $\pm$ 0.43 \\
      & \multicolumn{1}{>{\raggedright}m{1.6cm}}{SA\cite{souri2022sleeper}} & 88.31 $\pm$ 0.12 & -0.45 & 59.24 $\pm$ 1.41 & 65.15 $\pm$ 0.16 & -0.30 & 28.67 $\pm$ 0.64 & 57.38 $\pm$ 0.13 & -0.44 & 43.65 $\pm$ 0.78 \\
      & \multicolumn{1}{>{\raggedright}m{1.6cm}}{Ours} & 88.45 $\pm$ 0.08 & -0.31 & \cellcolor{red!25}82.29 $\pm$ 1.24 & 65.22 $\pm$ 0.13 & -0.23 & \cellcolor{red!25}75.25 $\pm$ 1.43 & 57.64 $\pm$ 0.78 & -0.18 & \cellcolor{red!25}57.39 $\pm$ 1.85 \\
      \hline

      \multicolumn{2}{>{\raggedright}m{2.0cm}}{\textbf{MobileNetV2}} &  &  &  &  &  &  &  &  & \\
      & \multicolumn{1}{>{\raggedright}m{1.6cm}}{Clean} & 90.32 $\pm$ 0.12 & 0 & 1.27 $\pm$ 0.34 & 64.18 $\pm$ 0.06 & 0 & 0.35 $\pm$ 0.12 & 51.92 $\pm$ 0.06 & 0 & 0.26 $\pm$ 0.08 \\
      & \multicolumn{1}{>{\raggedright}m{1.6cm}}{BadNet-c\cite{gu2019badnets}} & 87.75 $\pm$ 0.07 & -2.57 & 1.51 $\pm$ 0.24 & 61.45 $\pm$ 0.10 & -2.73 & 0.51 $\pm$ 0.06 & 50.24 $\pm$ 0.08 & -1.68 & 0.31 $\pm$ 0.05 \\
      & \multicolumn{1}{>{\raggedright}m{1.6cm}}{BadNet-d\cite{gu2019badnets}} & 88.24 $\pm$ 0.09 & -2.08 & 4.62 $\pm$ 0.14 & 60.97 $\pm$ 0.09 & -3.21 & 1.02 $\pm$ 0.07 & 50.14 $\pm$ 0.25 & -1.78 & 1.26 $\pm$ 0.13 \\
      & \multicolumn{1}{>{\raggedright}m{1.6cm}}{MIB\cite{hu2022membership}} & 88.19 $\pm$ 0.12 & -2.13 & 25.45 $\pm$ 0.23 & 61.05 $\pm$ 0.29 & -3.13 & 0.64 $\pm$ 0.25 & 50.46 $\pm$ 0.11 & -1.46 & 1.19 $\pm$ 0.36 \\
      & \multicolumn{1}{>{\raggedright}m{1.6cm}}{SA\cite{souri2022sleeper}} & 89.13 $\pm$ 0.12 & -1.19 & 36.40 $\pm$ 1.41 & 63.93 $\pm$ 0.16 & -0.25 & 30.67 $\pm$ 0.64 & 51.38 $\pm$ 0.13 & -0.54 & 34.65 $\pm$ 0.78 \\
      & \multicolumn{1}{>{\raggedright}m{1.6cm}}{Ours} & 89.78 $\pm$ 0.11 & -0.54 & \cellcolor{red!25}55.36 $\pm$ 2.42 & 64.05 $\pm$ 0.09 & -0.13 & \cellcolor{red!25}64.24 $\pm$ 2.37 & 51.74 $\pm$ 0.23 & -0.18 & \cellcolor{red!25}62.35 $\pm$ 1.72 \\
      \hline
      \multicolumn{2}{>{\raggedright}m{2.0cm}}{\textbf{Threshold}} & \multicolumn{3}{c}{\cellcolor{blue!25}23.35} & \multicolumn{3}{c}{\cellcolor{yellow!25}10.79} & \multicolumn{3}{c}{\cellcolor{green!25}9.94}\\
      \hline
    \end{tabular}
  }
  \captionsetup{size=large}
  \vspace{1em}
  \label{result}
\end{table*}

\paragraph{Metrics.}
In this paper, we use attack success rate (ASR), test accuracy, and clean accuracy drop \cite{xu2022poster} to evaluate the effectiveness of our proposed attack method. Our goal is to achieve a high attack success rate, high test accuracy, and low clean accuracy drop. Below, we elaborate on these three metrics.It is important to emphasize that the purpose of this paper is to employ backdoor techniques for data auditing, which differs from traditional Membership Inference Attacks (MIA). Traditional MIA typically uses the AUC score to assess privacy leakage, while backdoor techniques primarily focus on Attack Success Rate (ASR) and model accuracy.
\begin{itemize}
    \item \textbf{Attack Success Rate.}
The data owner can conduct black-box queries for $q$ poisoning test samples on the target model containing the backdoor and obtain its prediction results. We denote the attack success rate as $\alpha$, defined as follows:
    \begin{equation}
        \alpha =\frac{ \sum_{i=1}^{q} \mathbb{I} 
        \left ( f\left ( x_{p} ;\theta  \right )=y_{t} \right ) }{q} 
    \end{equation}  
    where $q$ represents the number of queries, $\mathbb{I}$ is the indicator function and $x_{p}$ denotes the poisoned sample containing the trigger. The value of $\alpha$ can be used as an estimate of the probability of success of the backdoor attack.
    \item \textbf{Test Accuracy.} We use test accuracy to measure the impact of backdoors on model performance. The ideal situation is to successfully implant the backdoor into the target model without affecting model performance.
    \item \textbf{Clean Accuracy Drop.} We also employ clean accuracy degradation (CAD) to assess the impact of the attack on the target model. CAD measures the disparity in classification accuracy between a clean target model and a target model containing a backdoor when evaluated on a clean test dataset.
\end{itemize}

\paragraph{Equipment.}
Our experiments were conducted on a deep learning server, which is equipped with an Intel(R) Xeon(R) Silver 4210 CPU @ 2.20GHz, 128GB RAM, and four NVIDIA GeForce RTX 3090 GPUs with 24GB of memory.

\subsection{Results}
\label{over1view1}

We compare our approach with an existing membership inference attack via a backdoor, namely MIB \cite{hu2022membership}. For performance comparison, we uniformly set the poisoning rate to \(0.1\%\). This is an extremely low setting. Additionally, we include the traditional backdoor attack, BadNets \cite{gu2019badnets}, in our comparison framework, dividing it into two forms: BadNets-c (clean label) and BadNets-d (dirty label). To underscore the efficiency of our proposed method, we also compare it with clean label-based backdoor attacks \cite{souri2022sleeper}.

Table \ref{result} reports the evaluation results of our proposed poisoning attack. Compared with other schemes, our method achieves the best attack results in most cases. Overall, we observe that our attack significantly increases the success rate of clean label backdoor attacks, but with a slight decrease in test accuracy. For example, for clean label backdoor attacks based on the VGG16 classifier on the CIFAR-10 dataset, the attack success rate increases from \(7.28\%\) to \(82.29\%\) on average, while the test accuracy only decreases by \(0.31\%\). The results show that the poisoned samples are similar to the clean samples, and the performance of the poisoned model decreases slightly on the test dataset. 

Another observation is that our clean-label backdoor attack method has better attack performance than dirty label attacks. A potential explanation for this phenomenon is that, due to the low poisoning rate and fewer poisoned samples containing triggers, the target model cannot learn the mapping relationship between triggers and labels very well in dirty label attacks. This also reflects the applicability of our method under low poisoning rates.

\section{Ablation Study }
\label{over1view1}

\begin{figure}[h]  
  \centering
  \includegraphics[width=\textwidth/2]{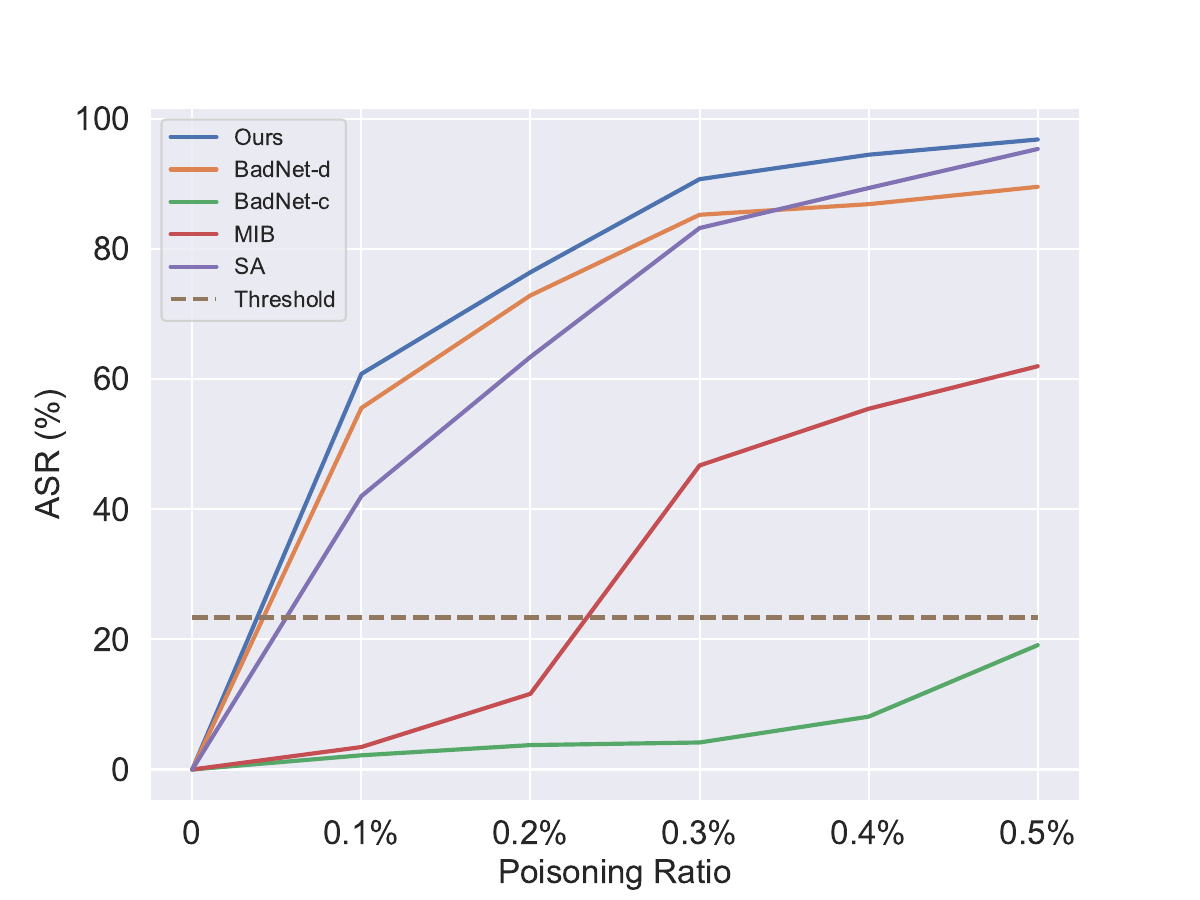}
  \caption{\textbf{ASR under different numbers of poisoned samples. We plot the curves on ResNet18 using the CIFAR-10 dataset.}}
  \label{fig:poison_amount}
\end{figure}



\subsection{Impact of Poisoning Rate}
\label{over1view2}

In this section, we examine the impact of the poisoning rate on attack effectiveness. Our experiments focused on attack performance at a poisoning rate of \(0.1\%\). This ablation study serves as a baseline to evaluate the efficacy of clean-label backdoor attacks. Figure \ref{fig:poison_amount} illustrates the observed attack performance. Notably, our findings indicate a significant enhancement in attack effectiveness with incremental increases in the poisoning rate. Our method consistently outperforms others by achieving a higher attack success rate, even at lower poisoning rates. For instance, with just 25 poisoned training samples, our attack performance reaches \(60.79\%\). Compared to poisoning rates reported in the literature \cite{chen2022amplifying,hu2022membership}, we explore attack performance under extremely low poisoning rates. When the poisoning rate is increased to \(0.5\%\), the attack performance escalates to \(96.81\%\), far exceeding the threshold.



\subsection{Impact of Noise }
\label{over1view3}

In our experiments, the perturbation radius significantly influences attack performance. We evaluate the effect of the perturbation radius on attack effectiveness. Figure \ref{fig:poison_size} shows the increase in attack success rate (ASR) as the perturbation range increases. Larger perturbations enable the poisoning of source class samples more effectively, allowing the shadow model to optimize the trigger by reducing the feature distance between source class samples and target samples. This results in improved attack performance. 

Intuitively, a larger perturbation radius can decrease the feature space distance between source class samples and target class samples. However, an excessively large perturbation radius may cause the original data samples to lose their semantics. To mitigate this issue, we maintain the same settings as in existing literature \cite{zeng2023narcissus, souri2022sleeper, chen2022amplifying}, setting the perturbation radius to \(\varepsilon = 16/255\), which is a standard and common configuration.

\subsection{Impact of Model }
\label{over1view4}


In our experiments, we relax the assumption that the shadow model must have the same architecture as the target model. Table \ref{shadow-target} presents the attack success rates (ASR) of various shadow models against the target model at a poisoning rate of \(0.1\%\). Notably, our attack achieves satisfactory ASR across most models. Interestingly, having identical architectures for the shadow and target models does not necessarily yield optimal ASR.

This finding suggests that attackers do not need to collect detailed information about the target model's architecture to maximize attack performance. Another noteworthy observation is that more complex models tend to provide better performance. Among the three architectures evaluated, VGG16, with the largest number of neurons, outperformed both ResNet18 and MobileNetV2. A plausible explanation for this is that more complex shadow models can better learn the differences between the source and target class labels during trigger generation, leading to the creation of more effective triggers. In particular, VGG16 consistently provided the best attack performance, regardless of the target model architecture.

\renewcommand{\arraystretch}{1.3}
\begin{table}[h]
\centering
\caption{\textbf{Experimental results for relaxing the shadow model assumptions. All results are on the CIFAR-100 dataset. The poisoning rate is $0.1\%$. The best ASRs are highlighted in bold.}}
\resizebox{0.9\columnwidth}{!}{
\begin{tabular}{c|ccc}
    \hline
    \diagbox[height=1.5cm,width=11em,dir=SE]{\strut \textbf{Target Models}}{\strut \textbf{Shadow Models}} & \textbf{ResNet18 } & \textbf{VGG16 } & \textbf{MobileNetV2 } \\
    \hline
    \textbf{ResNet18 } & 69.67 & \textbf{97.45} & 77.24 \\
    \textbf{VGG16 } & 68.28 & \textbf{75.29} & 15.36 \\
    \textbf{MobileNetV2 } & 63.14 & \textbf{95.42} & 66.24 \\
    \hline
\end{tabular}
}
\label{shadow-target}
\end{table}

\begin{figure*}[t]
    \centering
    \begin{subfigure}[b]{0.243\textwidth}
        \centering
        \includegraphics[width=\textwidth]{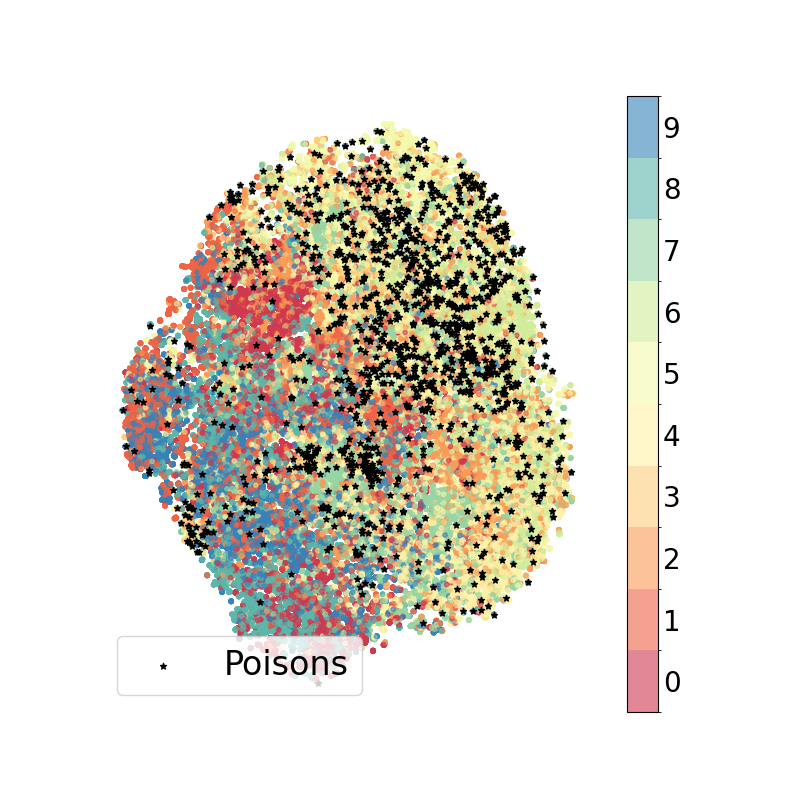}
        \caption{ }
        \label{fig:sub1}
    \end{subfigure}
    \hfill
    \begin{subfigure}[b]{0.243\textwidth}
        \centering
        \includegraphics[width=\textwidth]{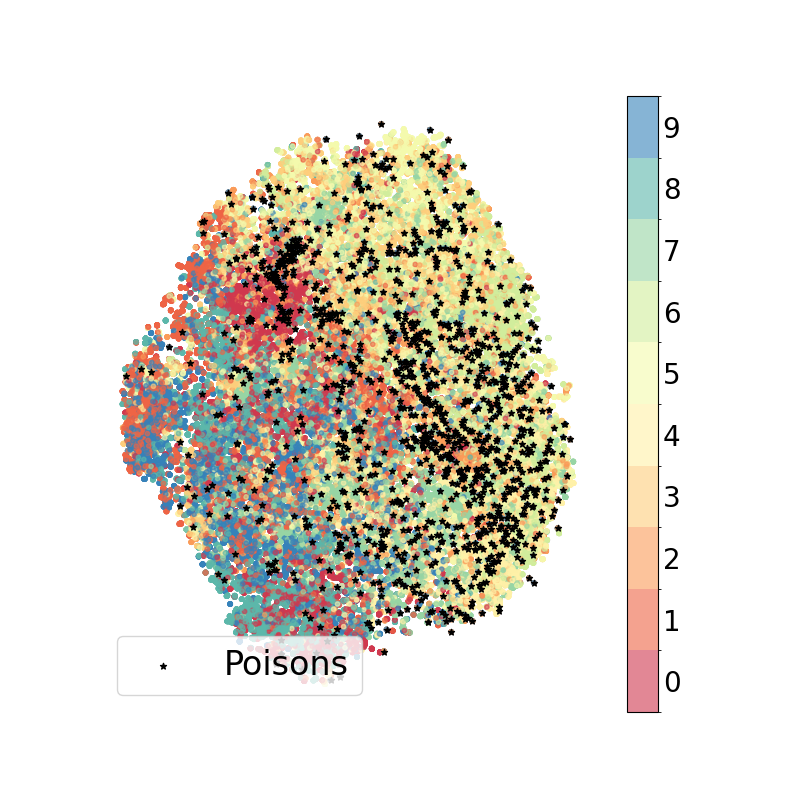}
        \caption{ }
        \label{fig:sub2}
    \end{subfigure}
    \begin{subfigure}[b]{0.243\textwidth}
        \centering
        \includegraphics[width=\textwidth]{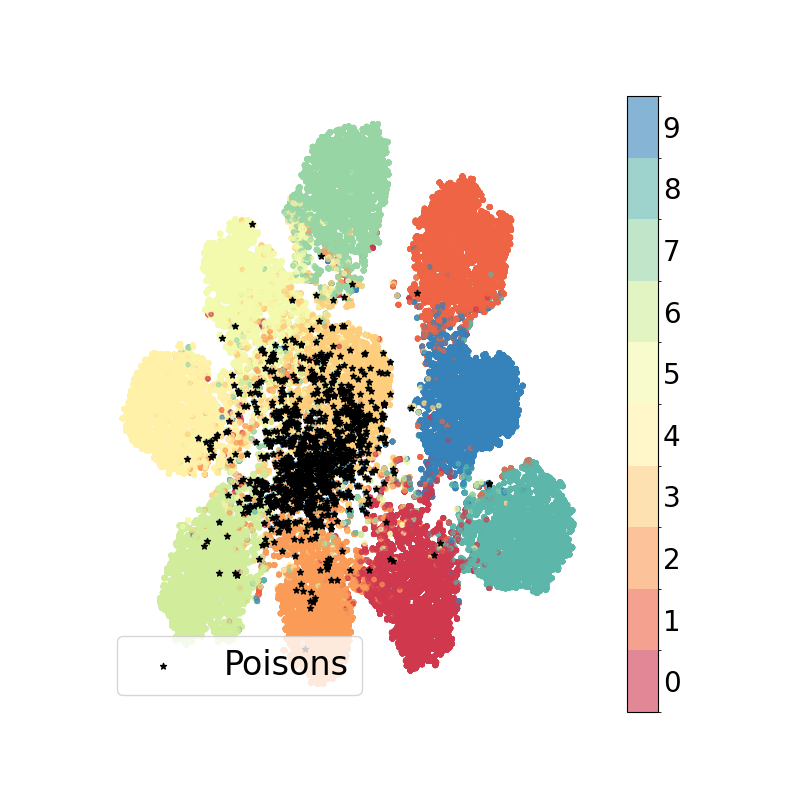}
        \caption{ }
        \label{fig:sub3}
    \end{subfigure}
    \hfill
    \begin{subfigure}[b]{0.243\textwidth}
        \centering
        \includegraphics[width=\textwidth]{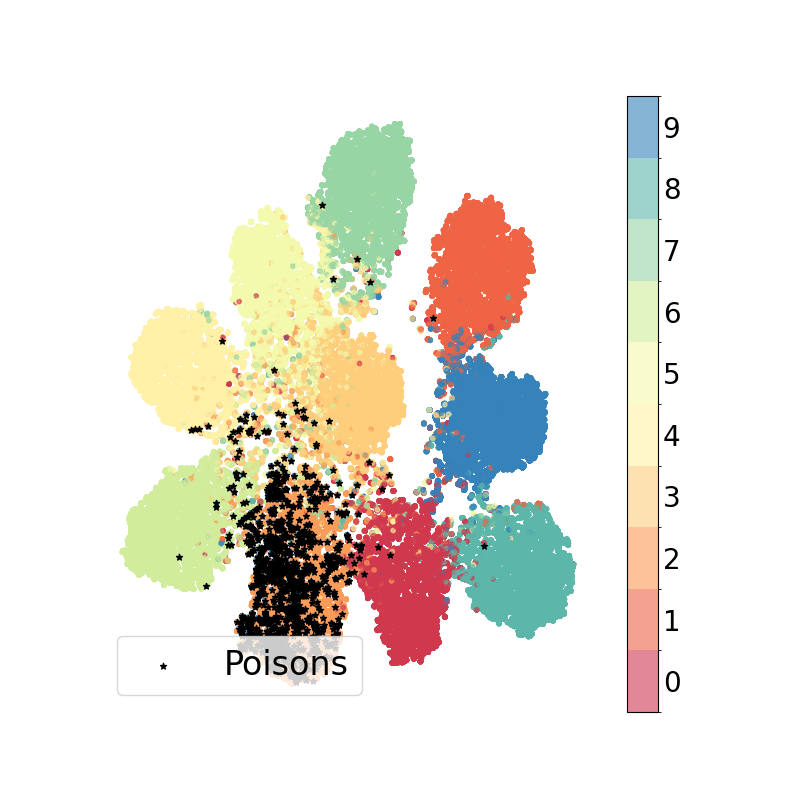}
        \caption{ }
        \label{fig:sub4}
    \end{subfigure}

    \vspace{12pt} 
    \caption{\textbf{Visualization of latent space features extracted from the ResNet18 feature extractor for different poisoning CIFAR-10 classifiers. The source category is 3 (cat) and the target category is 2 (bird). (a) Dirty label attack on untrained ResNet18. (b) Clean label attack on untrained ResNet18. (c) Clean label attack against trained clean ResNet18. (d) Clean label attack on trained poisoned ResNet18. Colored points are clean training samples, while dark star markers are poisoned samples.}}
    \label{fig:foursubfigures}
\end{figure*}

\begin{figure}[h]  
  \centering
  \includegraphics[width=\textwidth/2]{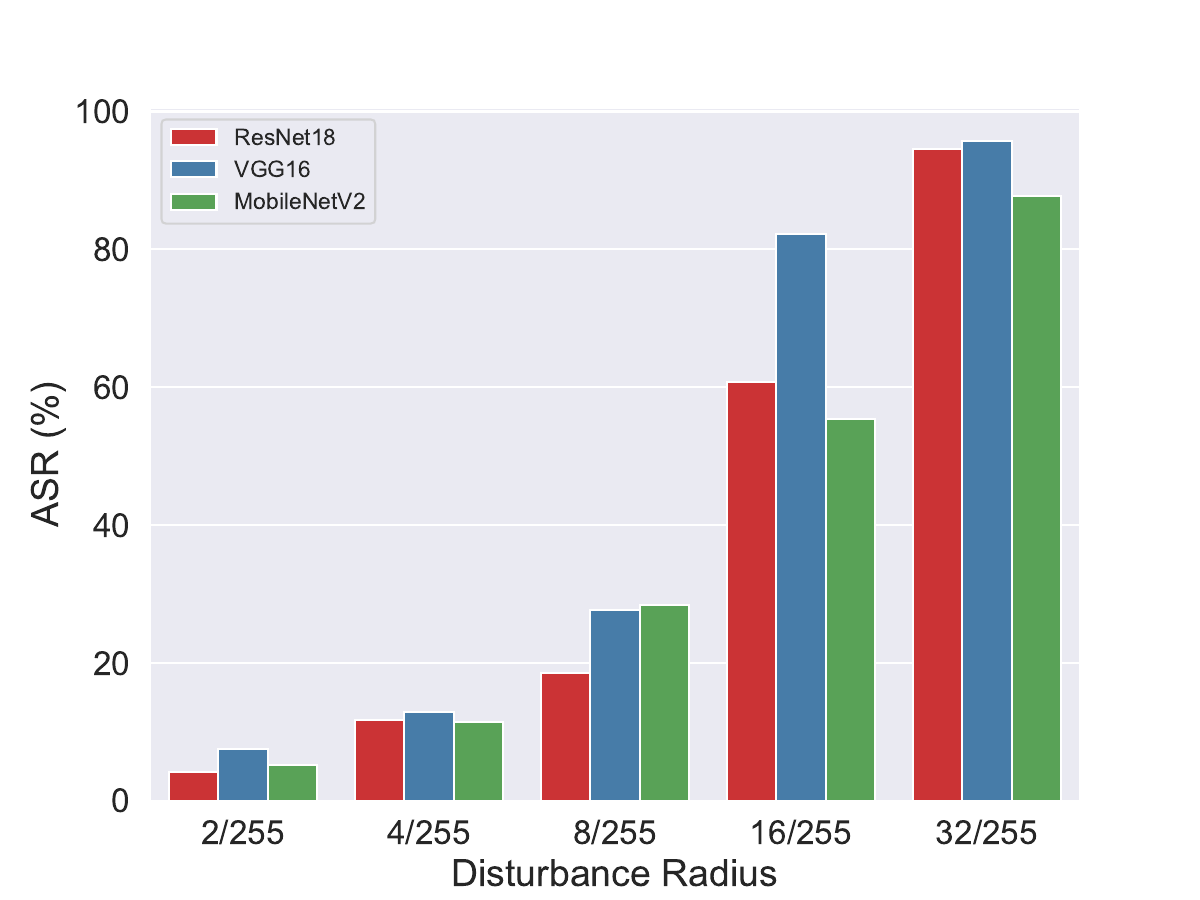}
  \caption{\textbf{ASR performance across varied perturbation radius on the CIFAR-10 dataset. The poisoning rate is $0.1\%$.}}
  \label{fig:poison_size}
\end{figure}

To understand this phenomenon, we investigated the behavior of the feature extractor under various conditions. We employed the t-SNE technique \cite{van2008visualizing} to visualize the latent features of clean and poisoned samples, as illustrated in Figure \ref{fig:foursubfigures}. For the clean label attack on the trained clean target model, the poisoned samples and the source class samples are closely clustered in the feature space (Figure \ref{fig:sub3}). This indicates that our attack method is ineffective against clean models. Conversely, when analyzing the clean label attack on the trained poisoned target model, we observe that the poisoned samples and the target class samples are also closely clustered in the feature space (Figure \ref{fig:sub4}). This demonstrates the effectiveness of our method, as the carefully designed triggers successfully shorten the feature space distance between the source class samples and the target samples.

\subsection{Impact of Label }
\label{over1view5}

\begin{figure}[h]  
  \centering
  \includegraphics[width=\textwidth/2]{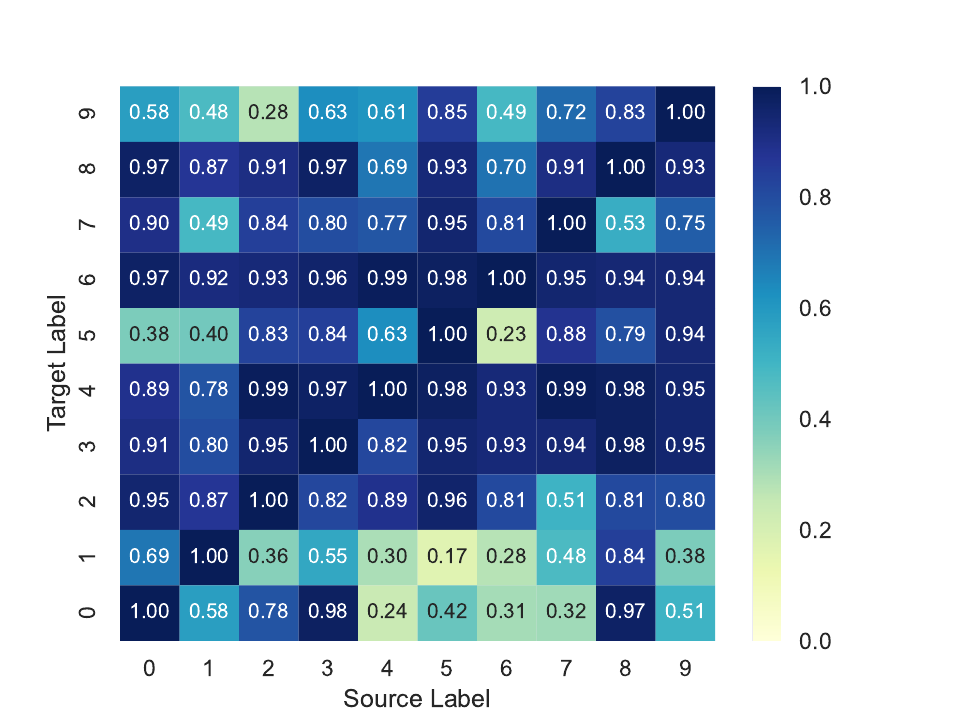}
  \caption{\textbf{The attack performance of ASR under different perturbation radius on the CIFAR-10 dataset. The poisoning rate is $0.1\%$.}}
  \label{fig:source_target}
\end{figure}
We investigated the influence of source and target label pairs on attack performance. Figure \ref{fig:source_target} illustrates the attack performance using various label combinations. To simplify interpretation, we set the attack success rate (ASR) to 1 when the source and target labels are identical. Our findings reveal consistently high attack performance across most label pairs. For example, when the source label is 7 and the target label is 4, the ASR can reach $99\%$, demonstrating the versatility of our attack method across different label pairs.

\subsection{Impact of the Number of Attackers }
\label{over1view5}

\begin{figure}[h]  
  \centering
  \includegraphics[width=\textwidth/2]{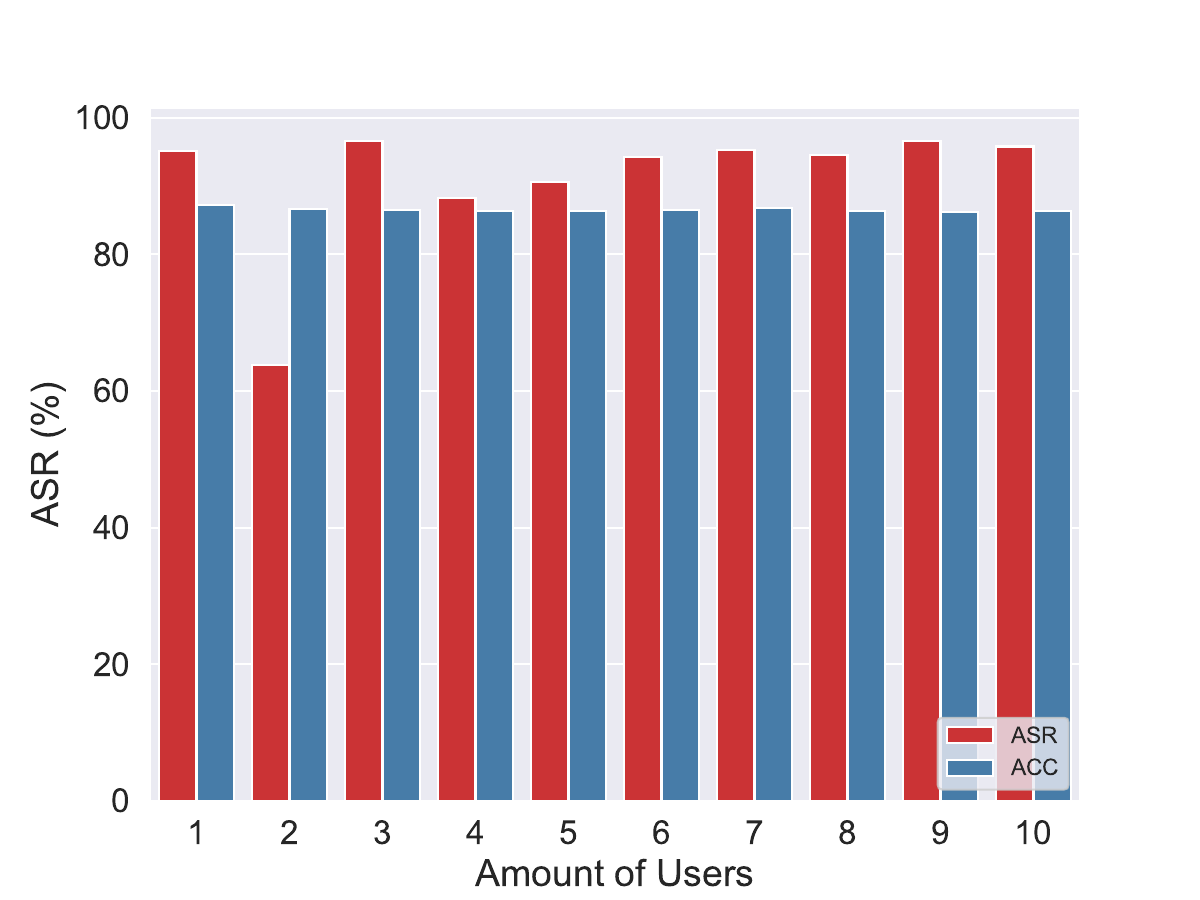}
  \caption{\textbf{Attack performance of ASR on the CIFAR-10 dataset with triggers from multiple users.}}
  \label{fig:users}
\end{figure}
We also explored the applicability of our approach when multiple users want to infer whether their private data is being used. It is worth exploring whether our carefully designed triggers can still work when multiple triggers are superimposed. Figure \ref{fig:users} shows the attack performance of our approach when triggers from multiple users are superimposed. We can observe that when multiple triggers are superimposed, our approach can achieve more than $99\%$ attack performance in most cases, and has little impact on the utility of the model. The reason is that when more triggers are added, more perturbations are added, and the greater the impact on model accuracy. Therefore, multiple users can work together to train a trigger for data auditing. When two users perform data auditing, a decline in the backdoor's Attack Success Rate (ASR) may occur. One potential reason is that their triggers interfere with each other, leading to a reduction in the effectiveness of the triggers.

\section{Related Work }
\label{sec:relwork}
\subsection{Membership Inference }
\label{sec:overview2}

As an emerging technique, membership inference attack aims to infer whether a specific sample $\left ( x ,y \right )$ belongs to the training data set $D_{train}$ of the target model. According to the attacker's capabilities, MIAs can be roughly divided into two categories: white-box attacks and black-box attacks:
 
\textbf{Black-box Membership Inference.} 
    In this case, the attacker distinguishes members and non-members only using model outputs \cite{shokri2017membership,hui2021practical,shokri2020exploiting}. There are generally two strategies in black-box settings: model-based attacks and metric-based attacks.

    \begin{itemize}
        \item \textbf{{Model-based Attacks:}} Shokri et al. \cite{shokri2017membership} introduced the first MIA against machine learning models, where the attacker has black-box access to the target model. The attacker builds multiple shadow models to mimic the target model, constructs a dataset of membership labels, and trains a binary classifier to predict membership status. However, this method requires extensive resources to train multiple shadow models with the same architecture as the target model, necessitating access to a shadow dataset with a distribution similar to that of the target's training set. To mitigate these issues, Salem et al. \cite{salem2019ml} proposed using a single shadow model, allowing for effective attacks while relaxing some adversarial assumptions.
        \item \textbf{{Metric-based Attacks:}} Song et al. \cite{song2021systematic} developed a metric-based attack where the attacker compares a calculated metric \(M\) (such as entropy) to a predefined threshold to infer membership. This method, however, is limited when the target model only provides predicted labels without prediction vectors. In contrast, our approach effectively utilizes predicted labels for membership inference. Bertran et al. \cite{bertran2024scalable} introduced a novel method that distinguishes between members and non-members using quantiles, eliminating the need for architectural knowledge and representing a true "black box" approach. Liu et al. \cite{liu2022membership} leverage the training process of the target model in their MIA, called TrajectoryMIA, utilizing knowledge distillation to extract membership information from loss records at various training epochs. These methods, while innovative, typically involve complex training processes and significant costs, particularly with knowledge distillation.
    \end{itemize}

\textbf{White-box Membership Inference.}  
In white-box settings, attackers gain access to model parameters \(\theta^{\ast}\) and potentially intermediate training information, such as gradients \(\frac{\partial \mathcal{L}}{\partial \theta}\) \cite{leino2020stolen, sablayrolles2019white}. This access can significantly enhance the effectiveness of membership inference attacks (MIAs), particularly in collaborative learning environments. For instance, Nasr et al. \cite{nasr2019comprehensive} introduced a white-box attack that targets privacy vulnerabilities inherent in the stochastic gradient descent algorithm used for training deep neural networks. Although many white-box attacks rely on model-based strategies, our focus here is on a black-box approach. It is important to note that leveraging the additional information available in white-box scenarios can be challenging in practical applications, making black-box attacks more relevant in many real-world contexts.

\subsection{Backdoor Techniques }
\label{sec:overview3}
In backdoor attacks, the primary objective is to inject poisoned samples containing triggers into the model's training set. These triggers embed a hidden backdoor in the model, enabling it to perform accurately on benign samples while altering its behavior when specific triggers are present. During testing, if a test sample contains the trigger, it activates the backdoor, causing the model to predict the target label associated with the trigger \cite{li2024backdoor}. Existing backdoor attacks can be broadly categorized into two main strategies: dirty-label attacks and clean-label attacks.

\textbf{Dirty-label Attacks.}  
Most traditional backdoor attacks rely on the attacker’s ability to control the labeling process, allowing them to modify the labels of poisoned samples \cite{gu2019badnets,li2020invisible,li2021invisible}. These attacks typically involve selecting clean samples from non-target classes, embedding backdoor triggers into the samples, and altering their labels to match the target class. Training on this poisoned dataset forces the model to learn the association between the trigger and the target label.

However, while these attacks can be highly effective, dirty label poisoning presents practical challenges in supervised machine learning settings. The altered labels often appear incorrect to human reviewers, making the poisoned samples easy to detect during manual inspection. Consequently, such poisoned samples are unlikely to remain in the dataset. In our comparative analysis, MIB \cite{hu2022membership} employs a dirty label attack, marking the first attempt to combine membership inference with backdoor techniques. Unfortunately, this approach suffers from two key issues: poisoned samples are readily identifiable, and the attack success rate is relatively low. To overcome these limitations, we propose using clean label backdoor attacks as a more effective alternative.

\textbf{Clean-label Attacks.}   
Clean label attacks aim to enhance the concealment of poisoned samples by ensuring that the input and its label appear consistent to human reviewers \cite{liu2020reflection,shafahi2018poison,barni2019new}. Turner et al. \cite{turner2019label} proposed two techniques for generating clean label poisoned samples. The first method embeds each sample into the latent space of a generative adversarial network (GAN) \cite{goodfellow2014generative} and interpolates the poisoned samples within the embedding of an incorrect class. The second method involves adding adversarial perturbations using optimization techniques to maximize the loss of a pre-trained model, thus making the model learn the trigger’s characteristics. However, Turner et al.'s methods often require a high poisoning ratio to establish the association between the trigger and the target label, which can be resource-intensive.

In contrast, Souri et al. \cite{souri2022sleeper} proposed a gradient matching objective to simplify the two-step optimization process, incorporating retraining and data selection functionalities. This approach, however, necessitates retraining the model, which leads to increased computational costs. Saha et al. \cite{saha2020hidden} introduced a technique using projected gradient descent (PGD) to optimize the trigger, creating poisoned images that are visually indistinguishable but are close to the target image in pixel space and close to the source image in feature space when patched with the trigger. Similarly, Zeng et al. \cite{zeng2023narcissus} developed a model-independent clean label backdoor attack that uses a surrogate model and out-of-distribution (POOD) data to generate triggers that point toward the target class.

While these methods are innovative, they all require access to real training samples for generating triggers, which is inconsistent with the assumptions in membership inference attacks where the actual training data is unknown. Our approach, in contrast, utilizes shadow models and shadow datasets to bypass the need for direct access to the training data, enabling effective attack performance. Compared to the aforementioned clean label methods, our technique requires minimal prior knowledge and achieves competitive attack success rates.

\section{Ethical and Privacy Considerations}
\label{sec:ethical}
In this section, we address the ethical and privacy considerations related to our proposed approach. Specifically, we introduce a novel method that leverages clean-label backdoors to audit membership inference. Although our approach advances the field by improving attack success rates and equipping data owners with tools to determine whether their data has been used in model training, it also raises critical ethical and privacy concerns.

\subsection{Dual-Use of MIA Techniques}
Membership inference attacks, though valuable for auditing potential privacy violations, come with dual-use implications. Techniques intended to safeguard individual privacy could, if misused, be employed maliciously to breach privacy by identifying individuals within a dataset. Given this adversarial potential, it is essential to frame these methods within a context of responsible use. Researchers and practitioners must ensure that MIA techniques are applied exclusively in ethical scenarios, such as privacy audits or regulatory compliance, and not for unauthorized exposure of data.

\subsection{Informed Consent and Data Ownership} 
The ability to infer membership in a training dataset can expose sensitive personal information, especially when dealing with datasets containing medical, financial, or other personally identifiable data. It is crucial that any application of MIAs is conducted with explicit consent from data owners. Data owners should be fully informed about how their data will be used and how the model will be audited, ensuring transparency throughout the process. Without such consent, applying MIAs could result in a violation of privacy and ethical research standards.

\subsection{Mitigating Potential Harms}
As with many adversarial techniques, there is a risk that the approach we propose could be misused to compromise privacy rather than protect it. To mitigate this, the research and development of MIAs should be guided by strict protocols, ensuring implementation in secure, controlled environments where data auditing is conducted with robust accountability measures. Furthermore, our clean-label backdoor approach should be applied with the explicit goal of enhancing security and privacy, allowing data owners to detect unauthorized model usage, rather than being used to facilitate attacks on machine learning models.

\subsection{Regulatory Compliance}
Current regulations, such as the General Data Protection Regulation (GDPR), emphasize the importance of user consent and data protection. Our work aligns with these regulatory goals by giving individuals a means to audit if their data has been used without permission. However, the potential misuse of these techniques also suggests that, if misused, they could conflict with data protection laws. Therefore, this work highlights the need for clear legal frameworks to govern the use of MIAs and backdoor techniques, ensuring they support data protection rather than undermine it.

\section{Conclusion}
\label{sec:conclusion}
In this paper, we present a novel approach termed Membership Inference via Clean-Label Backdoor, which effectively allows data owners to audit the usage of their data in model training. By strategically embedding triggers within clean-label samples, our method facilitates the detection of unauthorized data usage while remaining inconspicuous. Our findings demonstrate that this approach significantly enhances the accuracy of membership inference attacks on the target class, all while requiring only minimal black-box access to the target model and imposing limited adverse effects on its performance during inference. Notably, we achieve state-of-the-art results with just $0.1\%$ of poisoned samples labeled.

Through comprehensive evaluations, we explore various factors influencing attack performance, confirming the robustness and effectiveness of our method. Ultimately, our results underscore the urgent need to address the vulnerabilities of contemporary deep learning models in the context of data privacy. Future work will focus on refining techniques for crafting clean-label backdoors in a black-box setting, minimizing the need for underlying assumptions.





\end{document}